\documentclass[twocolumn,aps,prc,superscriptaddress,floatfix]{revtex4-1}
\usepackage{amssymb}
\usepackage{amsmath,bm}
\usepackage{graphicx}
\usepackage[normalem]{ulem}
\usepackage{bbding}
\usepackage{booktabs}
\usepackage[usenames, dvipsnames]{xcolor}
\usepackage{lipsum}
\usepackage{multirow}
\usepackage[colorlinks,
bookmarks=true,
linkcolor=blue,
urlcolor=blue,
anchorcolor=black,
citecolor=blue
]{hyperref}

\colorlet{darkblue}{blue!60!black}

\renewcommand{\sout}{\bgroup \color{red} \ULdepth=-.5ex \ULset}

\begin{document}

\title{Bayesian Inference of the Specific Shear and Bulk Viscosities of the Quark-Gluon Plasma at Crossover from $\phi$ and $\Omega$ Observables}



\author{Zhidong Yang}
\affiliation{School of Physics and Astronomy, Shanghai Key Laboratory for Particle Physics and Cosmology, and Key Laboratory for Particle Astrophysics and Cosmology (MOE), Shanghai Jiao Tong University, Shanghai 200240, China}


\author{Lie-Wen Chen}
\thanks{Corresponding author}
\email{lwchen@sjtu.edu.cn}

\affiliation{School of Physics and Astronomy, Shanghai Key Laboratory for Particle Physics and Cosmology, and Key Laboratory for Particle Astrophysics and Cosmology (MOE), Shanghai Jiao Tong University, Shanghai 200240, China}

\date{\today}

\begin{abstract}
Due to their weak final state interactions, the $\phi$ meson and $\Omega$ baryon provide unique probes of the properties of the quark-gluon plasma~(QGP) formed in relativistic heavy-ion collisions. Using the quark recombination model
with the quark phase-space information parameterized in a viscous blastwave, we study the transverse-momentum spectra and elliptic flows of $\phi$ and $\Omega$ in Au+Au collisions at $\sqrt{s_{\rm NN}} = 200$~GeV and Pb+Pb collisions at $\sqrt{s_{\rm NN}} = 2.76$~TeV.
The viscous blastwave includes non-equilibrium deformations of thermal distributions due to shear and bulk stresses and thus carries information on the specific shear viscosity $\eta/s$ and the specific bulk viscosity $\zeta/s$ of the QGP.
We perform a model-to-data comparison
with Bayesian inference and simultaneously obtain $\eta/s=(2.08^{+1.10}_{-1.09})/4\pi$ and $\zeta/s= 0.06^{+0.04}_{-0.04}$ at $90\%$ C.L. for the baryon-free QGP at crossover temperature of about $160$ MeV.
Our work
provides a novel approach to simultaneously determine the $\eta/s$ and $\zeta/s$ of the QGP at hadronization.
\end{abstract}


\maketitle

\section{Introduction}

A new state of matter that consists of deconfined quarks and gluons, called the quark-gluon plasma (QGP), is believed to have been created in relativistic heavy-ion collisions (HICs) at the Relativistic Heavy Ion Collider (RHIC) and the Large Hadron Collider (LHC). The QGP is expected to fill the early universe at about $10^{-6}$ s after the Big Bang. Studies show that the QGP behaves like a near-perfect fluid and has very small specific shear viscosity $\eta/s$~\cite{Romatschke:2007mq,Song:2007ux}, i.e., the ratio of shear viscosity $\eta$ to entropy density $s$, close to the universal lower bound $\eta/s=1/4\pi$ based on the AdS/CFT correspondence~\cite{Kovtun:2004de}. It turns out that the $\eta/s$ directly characterizes the behavior of QGP fluid and has significant impacts on the final observables of HICs. Studies also indicate the bulk viscosity ($\zeta$) has non-negligible effects on HIC observables~\cite{Ryu:2015vwa,Ryu:2017qzn}.

Determining the temperature and baryon density dependence of QGP's $\eta/s$ and $\zeta/s$ is of fundamental importance and remains a big challenge in the field~\cite{Shen:2020gef}.
For zero baryon density or bayron chemical potential ($\mu_B$=0), the first principle lattice QCD~(LQCD) predicts that the transition from hadrons to QGP is not a real phase transition but a smooth crossover at a pseudo-critical temperature $T_{\rm pc}\approx 160$~MeV~\cite{Bhattacharya:2014ara,Borsanyi:2020fev}.
Theoretical calculations on the temperature dependence of the viscosities of the baryon-free QGP have been widely explored in various approaches in the past years, including LQCD~\cite{Meyer:2007ic,Meyer:2009jp,Mages:2015rea,Astrakhantsev:2017nrs,Borsanyi:2018srz},
Functional Renormalization Group~(FRG) method~\cite{Christiansen:2014ypa}, T-matrix~\cite{Liu:2016ysz} and perturbative QCD~(pQCD)~\cite{Ghiglieri:2018dib} for the $\eta/s$, as well as LQCD~\cite{Karsch:2007jc,Meyer:2007dy,Astrakhantsev:2018oue} and holographic model~\cite{Rougemont:2017tlu} for the $\zeta/s$.
Overall, the theoretical predictions remain largely uncertain even at $T_{\rm pc}$ and the first principle calculation is still challenging \cite{Moore:2020pfu}.
Alternately,
many efforts have focused on constraining the viscosities of QGP with experimental observables.

Early works using viscous hydrodynamics have given semiquantitative estimates on the $\eta/s$ of QGP~\cite{Romatschke:2007mq, Schenke:2010rr,Gale:2012rq,Niemi:2015qia}. Usually they assumed a constant $\eta/s$ over the entire evolution and found $\eta/s=1\sim 2.5/(4\pi)$. Recently, results using multi-stage approaches which combine initial conditions, viscous hydrodynamics and hadronic transport are reported in Refs.~\cite{Bernhard:2015hxa,Bernhard:2016tnd,Bernhard:2019bmu,JETSCAPE:2020shq,JETSCAPE:2020mzn,Nijs:2020ors,Nijs:2020roc,Nijs:2022rme,Parkkila:2021tqq,Parkkila:2021yha,Heffernan:2023utr}. A variety of experimental observables, often focused on protons, kaons and pions, and a number of parameters are used in these works.
Although they used very similar methods, the obtained results tend to be different, depending on the details of the model and observables.
For example,
for the baryon-free QGP at $T_{\rm pc}$, JETSCAPE Collaboration reported $\eta/s \approx 1.6/(4\pi)$ and $\zeta/s \approx 0.09$~\cite{JETSCAPE:2020shq}, while Trajectum group reported $\eta/s \approx 0.9/(4\pi)$ and $\zeta/s \approx 0.004$ \cite{Nijs:2020ors}, displaying a significant discrepancy. Very recently Trajectum group updated their results to $\eta/s \approx 2.3/(4\pi)$ and $\zeta/s\approx 0.02$ at $T_{\rm pc}$ after considering the recent measured nucleon nucleon cross section \cite{Nijs:2022rme}. Meanwhile, they reported $\eta/s(T)$ decreases as temperature increases \cite{Nijs:2022rme}, which is opposite to the results of Duke group \cite{Bernhard:2019bmu}.

In this work, we present a novel approach to constrain the $\eta/s$ and $\zeta/s$ of the baryon-free QGP at $T_{\rm pc}$ by using
the measured transverse-momentum spectra and elliptic flows of $\phi$ and $\Omega$ in Au+Au collisions at $\sqrt{s_{\rm NN}} = 200$~GeV and Pb+Pb collisions at $\sqrt{s_{\rm NN}} = 2.76$~TeV where the QGP has a negligible baryon density and its transition to hadrons is a smooth crossover.
In our approach, hadrons are produced through quark recombination~\cite{Greco:2003xt,Greco:2003mm,Fries:2003vb,Fries:2003kq,Fries:2008hs,He:2010vw} with the phase-space distribution of quarks at hadronization parameterized in a viscous blastwave~\cite{Teaney:2003kp,Jaiswal:2015saa,Yang:2016rnw,Yang:2020oig,Yang:2022yxa} which includes non-equilibrium deformations of thermal distributions due to shear and bulk stresses. The viscous corrections on the QGP at hadronization are then imported into $\phi$ and $\Omega$ through the recombination process. Since the $\phi$ and $\Omega$ have weak hadronic interactions~\cite{Shor:1984ui}, they thus carry direct information of QGP at hadronization with negligible hadronic effects~\cite{Shor:1984ui,VanHecke:1998yu,Chen:2006vc,Chen:2008vr,Auvinen:2016uuv,Hwa:2016qtb,Ye:2017ooc,Pu:2018eei,Song:2019sez},
making it possible to determine the $\eta/s$ and $\zeta/s$ of the QGP at crossover from the $\phi$ and $\Omega$ observables.

\section{Model and methods}

\subsection{Theoretical model}

Quark recombination or coalescence models were first proposed to explain the baryon-over-meson enhancement and valence quark number scaling in RHIC Au+Au collisions \cite{Greco:2003xt,Greco:2003mm,Fries:2003vb,Fries:2003kq}. In these models, valence quarks are assumed to be abundant in the phase space and recombine into hadrons through quark recombination. The hadron formation process is usually assumed to be instantaneous and takes a very thin hypersurface ($\Delta\tau \approx 0$).  Following Refs.~\cite{Fries:2003kq,Fries:2003vb,Fries:2008hs}, the momentum distribution of baryons is given by
\begin{eqnarray}
E\frac{dN_B}{d^3\bf p}&=& C_B\int_\Sigma \frac{p^{\mu}\cdot d\mathbf{\sigma_\mu}}{(2\pi)^3}
\int_0^1 dx_1 dx_2 dx_3 \Phi_B(x_1,x_2,x_3)\nonumber\\ &&\times f_a({\bf r},x_1\mathbf{p})f_b({\bf r},x_2\mathbf{p})f_c({\bf r},x_3\mathbf{p}).\
\label{eq:baryon}
\end{eqnarray}
where $C_B$ is the spin degeneracy factor of a given baryon species, $\Sigma$ is the hypersurface of hadronization, $\varPhi_B$ is the effective wave function squared of baryons, $x_{1,2,3}$ are light cone coordinates defined as ${\bf p}_{1,2,3}=x_{1,2,3}\bf p$, and $f_{a,b,c}$ are the parton phase-space distributions. Here we make the assumption that the partons are essentially ultra-relativistic and traveling along the light cone, which should be valid for the LHC and the top RHIC energies. The $\varPhi_B$ is parameterized as polynomial \cite{Fries:2003kq}, and here we use a Gaussian ansatz
$\Phi_B\sim \exp (-\frac{(x_1-x_a)^2+(x_2-x_b)^2+(x_3-x_c)^2}{\sigma_B^2})\delta(x_1+x_2+x_3-1)$,
where $x_{a,b,c}=m_{1,2,3}/(m_1+m_2+m_3)$ are the peak values, and $m_{1,2,3}$ are the masses of the constitute partons. Similar expression can be derived for mesons.

The quark phase-space distribution is parameterized in a viscous blastwave~\cite{Yang:2016rnw,Yang:2020oig,Yang:2022yxa}, based on Retiere and Lisa (RL) blastwave \cite{Retiere:2003kf}. The quark distribution is given by
\begin{equation}
f(r,p) = f_{0}(r,p) + \delta f_{\rm {shear}}(r,p)+\delta f_{\rm{bulk}}(r,p)
\label{eq:fdist}
\end{equation}
where $f_0$ is the equilibrium Bose/Fermi distribution, $\delta f_{\rm {shear}}$ and $\delta f_{\rm{bulk}}$ are corrections from the shear and bulk viscosities, respectively. For the shear viscous corrections, we use the Grad's method \cite{Song:2007ux, Damodaran:2017ior}
\begin{equation}
\delta f_{\rm {shear}} = \frac{1}{2s} \frac{p^\mu p^\nu}{T^3}\pi^{\mu\nu} f_{0}(1\pm f_{0})
\label{eq:shear}
\end{equation}
where $\pi^{\mu\nu}$ is the shear stress tensor and $+(-)$ for bosons (fermions). In the Navier-Stokes approximation $\pi^{\mu\nu}=2\eta \sigma^{\mu\nu}$ where $\sigma^{\mu\nu}$ is the shear gradient tensor defined as $
\sigma^{\mu\nu} = \frac{1}{2} \left( \nabla^\mu u^\nu + \nabla^\nu u^\mu \right) - \frac{1}{3} \Delta^{\mu\nu} \nabla_\lambda u^\lambda$ with flow field $u^\mu$, $\nabla^\mu = \Delta^{\mu\nu} \partial_\nu$ and $\Delta^{\mu\nu} = g^{\mu\nu} - u^\mu u^\nu$. For the bulk viscous corrections, we use 14-moment approximation \cite{Denicol:2014vaa,Paquet:2015lta}
\begin{equation}
\delta f_{\rm bulk}=- f_0(1\pm f_0) \Pi \frac{\tau_\Pi}{\zeta} \left[ \frac{1}{3} \frac{m^2}{T} \frac{1}{p^\mu u_\mu}+\frac{p^\mu u_\mu}{T} \left(c_s^2-\frac{1}{3} \right) \right]
\label{eq:bulk1}
\end{equation}
where $\Pi$ is the bulk viscous pressure and $\tau_\Pi$ is the bulk relaxation time. At the first order approximation, one has $\Pi=-\zeta \partial_\mu u^\mu$. The expression for $\frac{\tau_\Pi}{\zeta}$ is given in \cite{Denicol:2014vaa}.

Now we present the blastwave parameterization for the flow field $u^\mu$. In the following, $R_{x,y}$ are the semi axes of the fireball at freeze-out, $\rho=\sqrt{x^2/R_x^2+y^2/R_y^2}$ is the reduced radius, $\eta_s = \frac{1}{2}\log\frac{t+z}{t-z}$ is the space-time rapidity. The hypersurface is assumed to be constant $\tau=\sqrt{t^2-z^2}$. The flow field is parameterized as
\begin{eqnarray}
	u^{\mu}&=&(\cosh\eta_s\cosh\eta_T,\sinh\eta_T\cos\phi_b, \nonumber\\ && \qquad \sinh\eta_T\sin\phi_b,\sinh\eta_s\cosh\eta_T)
	\label{eq:flow}
\end{eqnarray}
where $\eta_T$ is the transverse flow rapidity and $\phi_b$ is the azimuthal angle of $u^\mu$ in the transverse plane. Here we assume boost invariance and set longitudinal flow rapidity to be equal to the space-time rapidity $\eta_s$. $\eta_T$ is given by the transverse velocity $v_T=\tanh\eta_T$ with
\begin{equation}
v_T=\rho^n (\alpha_0+\alpha_2\cos2\phi_b)
\end{equation}
where $\alpha_0$ is the average surface velocity, $\alpha_2$ is an elliptic deformation of the flow field and $n$ is a power term. In this work, we use a linear expression and set $n=1$. We choose the transverse flow vector perpendicular to the elliptic surface at $\rho=1$, i.e., $\tan \phi_b = R_x^2/R_y^2 \tan \phi$, where $\phi=\arctan y/x$ is the azimuthal angle of the position. In terms of the geometric parameters, it turns out the ratio $R_y/R_x$ has a large influence on elliptical flow, so we choose $R_y/R_x$ as a fit parameter and constrain $R_x$, $R_y$ and $\tau$ by fitting the ratio $R_y/R_x$ and by adding the simple geometric estimate $R_x \approx (R_0-b/2)+ \tau c_\tau (\alpha_0+\alpha_2)$ where $R_0$ is the radius of the colliding nucleus, $b$ is the impact parameter and $c_\tau = \bar\alpha_0 / \alpha_0$ relates the time-averaged surface velocity. In this work we use $R_0=7.1$ fm (7.0 fm) for Pb (Au) based on formula in \cite{Nerlo-Pomorska:1994dhg} and $c_\tau=0.65$. The values of $b$ for each centrality bin are taken from Glauber Monte Carlo simulations used by related experiments \cite{ALICE:2013hur,HADES:2017def}.

It should be noted that
the viscous blastwave parameterizes the flow field and freeze-out hypersurface with a simple ansatza, which can be regarded as an approximate snapshot of a viscous hydrodynamic system at a fixed time~\cite{Teaney:2003kp,Jaiswal:2015saa,Yang:2016rnw,Yang:2020oig,Yang:2022yxa}. Therefore, the viscous blastwave carries information on the viscosities of the fluid at a fixed time, e.g., the QGP at hadronization in our present work.

\subsection{Experimental data and fit parameters}

To compare our model to experiment, we utilize the transverse-momentum~($p_T$) spectra and elliptic flows $v_2$ of $\phi$ and $\Omega$ as our observables. The data are taken from the STAR Collaboration for Au+Au collisions at $\sqrt{s_{\rm NN}} = 200$~GeV~\cite{STAR:2006egk,STAR:2007mum,STAR:2015gge} and the ALICE Collaboration for Pb+Pb collisions at $\sqrt{s_{\rm NN}} = 2.76$~TeV~\cite{ALICE:2013xmt,ALICE:2014wao,ALICE:2017ban} at mid-rapidity.
For $v_2$ of $\phi$ and $\Omega$,
we use the data in centralities 0-30\%, 30-80\% from STAR and 10-20\%, 20-30\%, 30-40\%, 40-50\% from ALICE. Due to the limit of available data, the centrality bins for $p_T$ spectra are slightly different from that of $v_2$. For the $\phi$ spectra, we use the data in 10-20\%, 40-50\% from STAR and 10-20\%, 20-30\%, 30-40\%, 40-50\% from ALICE. For the $\Omega$ spectra, we use the data in 0-5\% and 40-60\% from STAR and 10-20\%, 20-40\%, 40-60\% from ALICE. When the centrality of spectra are not matched to that of $v_2$, we choose the spectra that belongs to the neighbor centrality.

The fit ranges are $p_T <$ 1.8 GeV/$c$ for $\phi$ and $p_T<$ 3.2 GeV/$c$ for $\Omega$. We do a combined analysis for both STAR and ALICE data and have 116 data points in total. We note that the correction $\delta f(r,p)$ is small for both $\phi$ and $\Omega$, i..e., less than $20\%$ of $f_0$ for the majority of points (very few points up to $40\%$ of $f_0$), which is much smaller than the usually adopted upper bound $\sim 1$ \cite{JETSCAPE:2020mzn,Teaney:2003kp}, guaranteeing the applicability of the viscous corrections.

The parameters entering the model are ($\tau$, $T$, $\alpha_0$, $\alpha_2$, $R_y/R_x$, $\eta/s$, $\zeta/s$) from blastwave and ($\sigma_B$, $\sigma_M$) from the quark recombination. Here $\sigma_B$($\sigma_M$) is the wave function width for baryons (mesons). We find the values of $\sigma_{M,B}$ mainly affect the particle yields and have little impact on $v_2$. To describe the yields of $\phi$ and $\Omega$, we set $\sigma_M=0.3$, $\sigma_B=0.09$. We note that our conclusion changes little by varying the values of $\sigma_M$ and $\sigma_B$. The freeze-out or recombination hypersurface is related to the fireball volume and can be determined from the particle yield.

For the other constants, we use sound speed squared $c^2_s=0.15$ (see Eq.\ (\ref{eq:bulk1})) for the baryon-free QGP at $T_{\rm pc}$ \cite{Huovinen:2009yb}, quark mass $m_s=0.5$ GeV, spin degeneracy factor $C_M=3$ for $\mathrm{\phi}$ and $C_B=4$ for $\Omega^-$ (or $\bar{\Omega}^+$). To fit the yield of $\phi$ and $\Omega$ as well as determine the value of freeze-out time $\tau$, we introduce a fugacity factor $\gamma_{s,\bar{s}}$ for strange quarks and set $\gamma_{s,\bar{s}}=0.8$ for all centrality bins which is close to the value found in \cite{Pu:2018eei}. Now the parameters left in our model are ($T, \alpha_0, \alpha_2, R_y/R_x, \eta/s, \zeta/s$). For each centrality bin of STAR (0-30\%, 30-80\%) and ALICE (10-20\%, 20-30\%, 30-40\%, 40-50\%), the fluid has unique values for $\alpha_0$, $\alpha_2$ and $R_y/R_x$ with the same shared values for $T$, $\eta/s$ and $\zeta/s$, and thus we have 21 parameters.

\begin{figure}[tbh]
	\centering
	\includegraphics[width=0.95\linewidth]{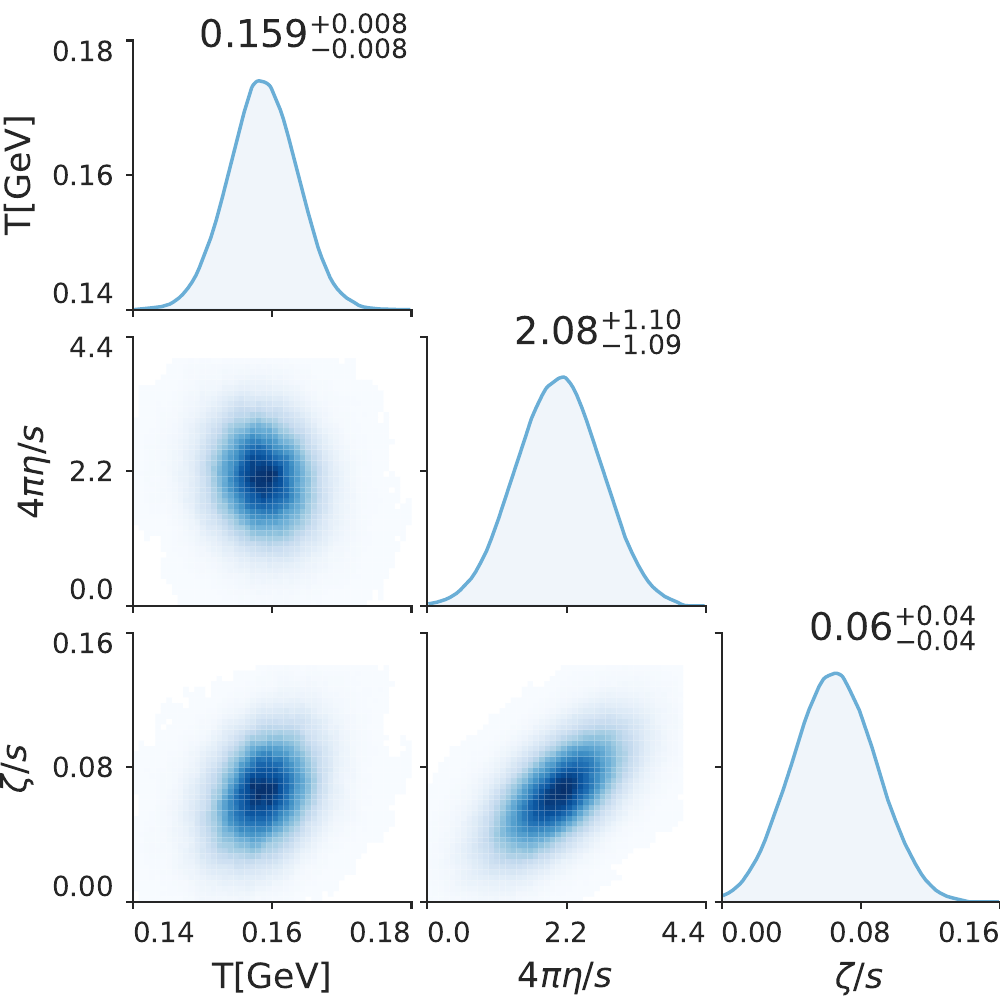}
	\caption{\label{fig:posterior}
		Posterior distributions and correlations of selected parameters: $T$, $\eta/s$ and $\zeta/s$. The numbers indicate the median values with the 90\%-credibility range.}
\end{figure}

\subsection{Bayesian method}

To determine the above parameters, we use the Bayesian analysis package from the Models and Data Analysis Initiative (MADAI) project \cite{MADAI:2013,Bernhard:2016tnd}. The MADAI package includes a Gaussian process emulator and a Bayesian analysis tool. According to Bayes' theorem, for model parameters $\rm x=(x_1,x_2,x_3,...)$ and experimental observables $\rm y=(y_1,y_2,y_3,...)$, the probability for the true parameters $\rm x_\star$ is
\begin{equation}
P({\rm x}_\star|X,Y,{\rm y}_{\rm {exp}}) \propto P(X,Y,{\rm y}_{\rm {exp}}|{\rm x}_\star) P({\rm x}_\star).
\label{eq:bayes}
\end{equation}
The left-hand side is the \emph{posterior} probability of ${\rm x}_\star$ given the design ($X$, $Y$) and the experimental data ${\rm y}_{\rm {exp}}$. On the right-hand side, $P({\rm x}_\star)$ is the \emph{prior} probability and $P(X,Y,{\rm y}_{\rm {exp}}|{\rm x}_\star)$ is the likelihood, i.\ e.\ , the probability of the model describing the data ${\rm y}_{\rm {exp}}$ at ${\rm x}_\star$, given by
\begin{equation}
P(X,Y,{\rm y}_{\rm {exp}}|{\rm x}_\star)\propto\exp (
-\frac{1}{2} \Delta {\rm y}^\top \Lambda^{-1} \Delta {\rm y})
\end{equation}
where $\Delta {\rm y}={\rm y}_\star - {\rm y}_{\rm {exp}}$ is the difference between the measurement and the prediction, and $\Lambda$ is the covariance matrix including the experimental and model uncertainties. More details can be found in  \cite{Bernhard:2016tnd,JETSCAPE:2020mzn}.

All model parameters have \emph{uniform} prior distribution except for the recombination temperature $T$, which has a \emph{Gaussian} prior. For example, we use prior ranges 0-4.4/($4\pi$) for $\eta/s$, 0-0.16 for $\zeta/s$ and 0.72-0.8$c$ for $\alpha_0$, 0.02-0.036$c$ for $\alpha_2$, 1.1-1.18 for $R_y/R_x$ in 0-30\% centrality bin of STAR. For $T$, we use Gaussian distribution with a mean value of 160 MeV and a standard deviation of 5 MeV to reflect our prior knowledge of quark recombination temperature of a continuous crossover at $T_{\rm pc} = 158\pm 15$ MeV for zero net baryon density~\cite{Borsanyi:2020fev}.

The Bayesian analysis package works as follows. First, we set prior ranges for each parameter and generate a set of training points in the parameter space. Second, we calculate all fitted observables at each training point. The package then builds a Gaussian process emulator, which can estimate the observables for random parameter values. Finally a Markov Chain Monte Carlo (MCMC) provides a likelihood analysis and gives the maximum likelihood or best-fit parameters. Here we have 21 parameters and use N=500 training points. To verify the Bayesian analysis works properly, we perform a validation for the Bayesian inference framework and find the Bayesian framework correctly reproduces model parameters within reasonable uncertainties. The likelihood analysis has used ${\rm N}_\star=10^7$ predicted points to search for the best-fit parameters, which is sufficient for MCMC to converge.

\section{Results and discussions}

Using the data and prior parameter ranges, we perform a model-to-data comparison with MADAI package. Fig.\ \ref{fig:posterior} shows the univariate posterior distributions for the recombination temperature $T$, $\eta/s$ and $\zeta/s$. We obtain $T=159\pm 8$ MeV, $\eta/s=(2.08^{+1.10}_{-1.09})/4\pi$ and $\zeta/s= 0.06^{+0.04}_{-0.04}$ at $90\%$ confidence level~(C.L.). Fig.\ \ref{fig:posterior} also shows the correlations between $T$, $\eta/s$ and $\zeta/s$. We find there is a strong correlation between $\eta/s$ and $\zeta/s$. This can be qualitatively understood by the different effects of $\eta/s$ and $\zeta/s$ on the particle momentum distribution, i.e., the $\eta/s$ decreases the azimuthal momentum asymmetry of particles, while the $\zeta/s$ reduces their momenta isotropically~\cite{JETSCAPE:2020shq}. As a result, the $\zeta/s$ tends to increase $v_2$ while the $\eta/s$ prefers to reduces $v_2$ at higher $p_T$. We also find there is a slight anti-correlation between $T$ and $\eta/s$, and a moderate correlation between $T$ and $\zeta/s$.

\begin{table}[tb]
	\caption{\label{tab:recombrt} The  best-fit values of the viscous blastwave parameters for different centrality bins with $T=159$ MeV, $\eta/s = 2.08/(4\pi)$ and $\zeta/s = 0.06$.}
	\centering
	\begin{tabular}{|lllll|} 
		\hline
		Centrality & $\tau$ (fm/$c$) & $\alpha_0(c)$ &$\alpha_2(c)$  & $R_y/R_x$  \\
		\hline
		\multicolumn{5}{|l|} {Au+Au 200 GeV}  \\
		\hline
		0-30\% & 7.8  & 0.76 & 0.027  & 1.13  \\
		30-80\% & 5.1   & 0.70 & 0.039  & 1.31  \\
		\hline
		\multicolumn{5}{|l|}{Pb+Pb 2.76 TeV}  \\
		\hline
		10-20\% & 9.4  & 0.84 & 0.028  & 1.18  \\
		20-30\% & 9   & 0.81 & 0.037  & 1.18  \\
		30-40\% & 7.8    & 0.81 & 0.036  & 1.25 \\
		40-50\% & 6.8 & 0.79 & 0.031  & 1.36   \\
		\hline
	\end{tabular}	
\end{table}

\begin{figure}[tbh]
	\centering
	\includegraphics[width=0.95\linewidth]{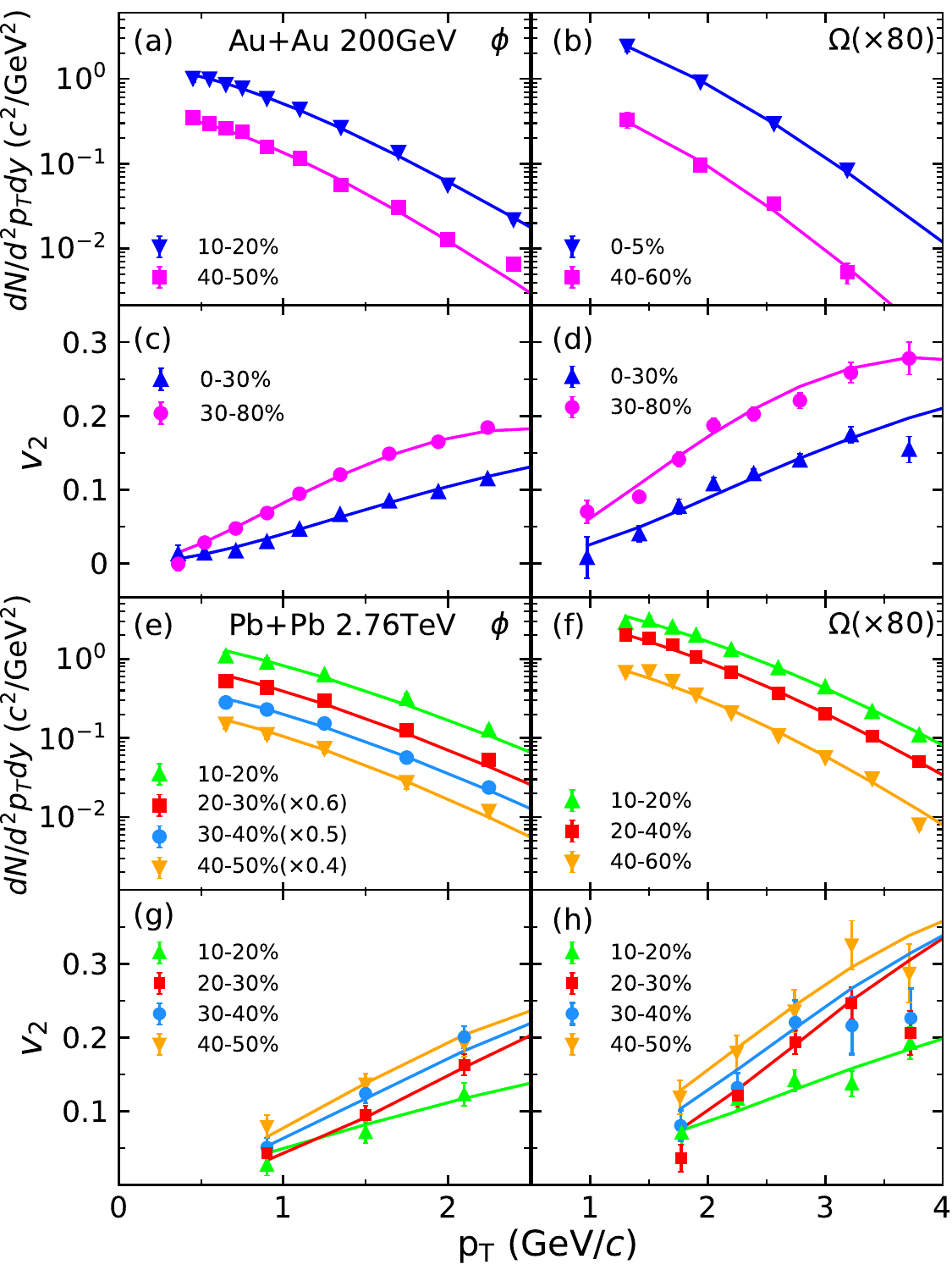}
	\caption{\label{fig:qrspv2}
Transverse-momentum spectra and elliptic flows $v_2$ of $\phi$ and $\Omega^-$+$\bar{\Omega}^+$ at mid-rapidity in Au+Au collisions at $\sqrt{s_{\rm NN}} = 200$~GeV and Pb+Pb collisions at $\sqrt{s_{\rm NN}} = 2.76$~TeV from the recombination model using the best-fit parameters given by Baysian analysis. The data from STAR~\cite{STAR:2006egk,STAR:2007mum,STAR:2015gge} and ALICE~\cite{ALICE:2013xmt,ALICE:2014wao,ALICE:2017ban} are included for comparison.}
\end{figure}

For the other parameters, we show the best-fit values, which are defined as the mean value given by maximum likelihood analysis, as given in Tab.\ \ref{tab:recombrt}. Using the best-fit parameters, we calculate the $p_T$ spectra and $v_2$ of $\phi$ and $\Omega$ and compare them with data. Fig.~\ref{fig:qrspv2} shows the results of $p_T$ spectra and $v_2$ in different centrality bins for Au+Au collisions at $\sqrt{s_{\rm NN}} = 200$~GeV and Pb+Pb at $\sqrt{s_{\rm NN}} = 2.76$~TeV, respectively. As expected, our calculations describe the data rather well.

\begin{figure}[tbh]
		\centering
		\includegraphics[width=0.95\linewidth]{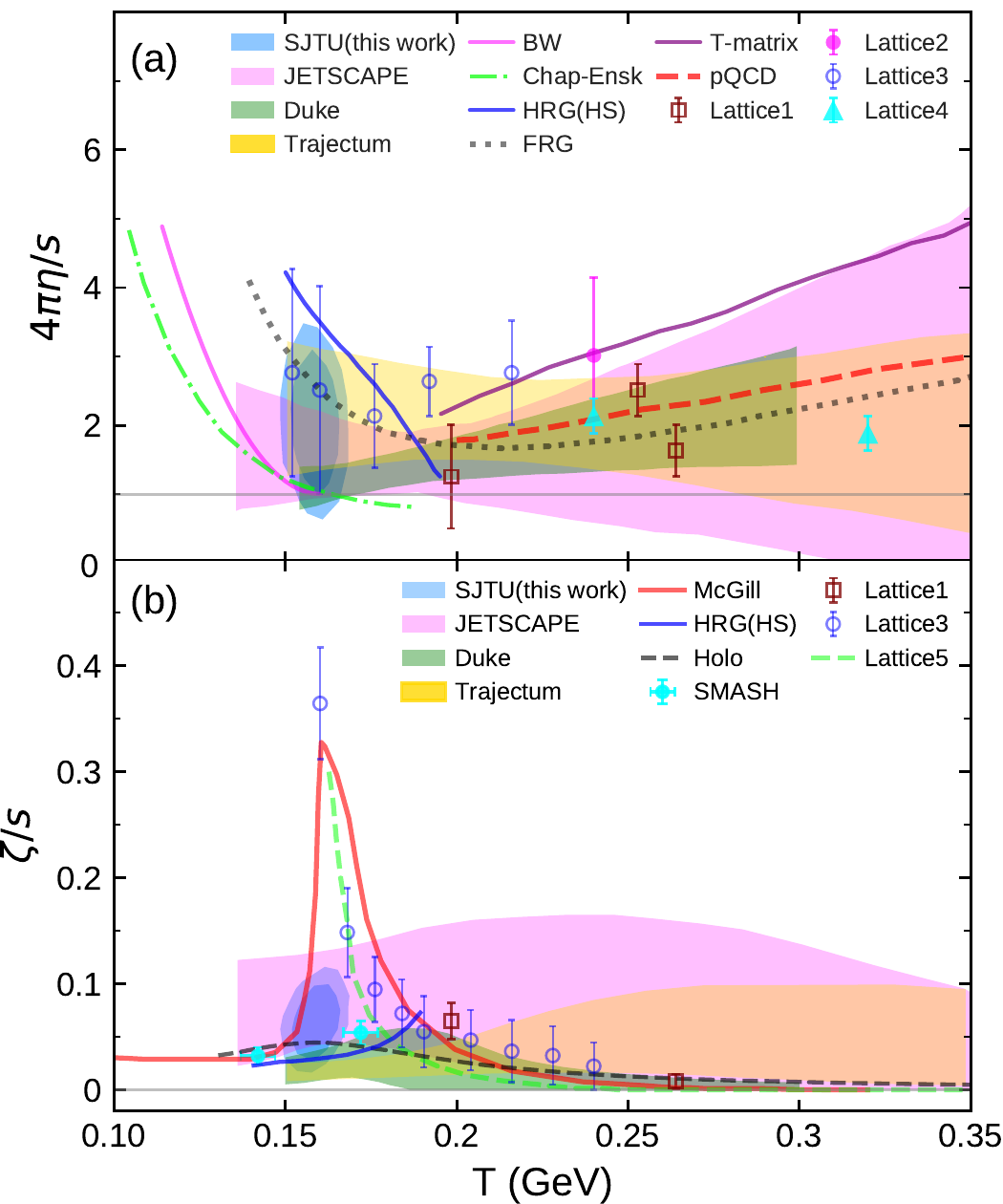}
        \caption{\label{fig:region}
	Temperature dependence of $\eta/s$~(a) and $\zeta/s$~(b) for QGP/hadronic matter at $\mu_B$=0 (see text for details).}
\end{figure}

Figure~\ref{fig:region} shows our Bayesian inference of the $\eta/s$ and $\zeta/s$ for the baryon-free QGP at $T_{\rm pc}\approx 160$ MeV~(with 68.3\% and 90\% C.L., indicated by SJTU).
For comparison,
we include in Fig.~\ref{fig:region}(a) the temperature dependence of $\eta/s$ for the baryon-free QGP/hadronic matter from other approaches, i.e.,
the multi-stage methods (with 90\% C.L.) from Duke~\cite{Bernhard:2019bmu}, JETSCAPE~\cite{JETSCAPE:2020shq} and Trajectum~\cite{Nijs:2022rme},
the viscous blastwave~(BW)~\cite{Yang:2022yxa}, Chapman-Enskog~(Chap-Ensk) method~\cite{Dash:2019zwq}, hadron resonance gas~(HRG) model with Hagedorn states~(HS)~\cite{Noronha-Hostler:2008kkf}, FRG~\cite{Christiansen:2014ypa}, LQCD~(Lattice1~\cite{Meyer:2007ic,Meyer:2009jp}, Lattice2~\cite{Mages:2015rea}, Lattice3~\cite{Astrakhantsev:2017nrs} and Lattice4~\cite{Borsanyi:2018srz}, with 68.3\% C.L.), T-matrix~\cite{Liu:2016ysz}, and next-to-leading order pQCD~\cite{Ghiglieri:2018dib}.
Similarly,
Fig.~\ref{fig:region}(b)
includes the corresponding results for $\zeta/s$ from
the multi-stage methods from Duke~\cite{Bernhard:2019bmu}, JETSCAPE~\cite{JETSCAPE:2020shq} and Trajectum~\cite{Nijs:2022rme},
LQCD (Lattice1~\cite{Meyer:2007dy}, Lattice3~\cite{Astrakhantsev:2018oue} and Lattice5~\cite{Karsch:2007jc}, with 68.3\% C.L.), hybrid model (McGill)~\cite {Ryu:2015vwa}, holographic model~(Holo)~\cite{Rougemont:2017tlu}, HRG model with HS~\cite{Noronha-Hostler:2008kkf} and SMASH transport model~\cite{Rose:2020lfc}.

It is interesting to see from Fig.~\ref{fig:region}(a) that different methods give roughly consistent results for $\eta/s$, indicating a trend that the $\eta/s$ approaches to a minimum at the crossover temperature $T_{\rm pc}$, with a rather steep rise towards lower temperatures and a slow rise towards higher temperatures. The behavior with minimum $\eta/s$ at $T_{\rm pc}$ is also observed in recent work based on 2 + 1D hydrodynamical model with the Eskola-Kajantie-Ruuskanen-Tuominen (EKRT) initial state~\cite{Auvinen:2020mpc} and the quasiparticle
model prediction~\cite{Mykhaylova:2019wci}.
We note our present constraint on the $\eta/s$ at $T_{\rm pc}$ is compatible with those of Duke, JETSCAPE and Trajectum from the multi-stage methods, as well as the Chap-Ensk, BW, FRG and Lattice3.

On the other side, one sees from Fig.~\ref{fig:region}(b) that the results of $\zeta/s$  exhibit significant discrepancy around $T_{\rm pc}$. For example, at $T_{\rm pc}$,
a rather large value of $\zeta/s \approx 0.35$ is obtained from the LQCD (Lattice3 and Lattice5) and McGill group, $\zeta/s \approx 0.09$ from JETSCAPE, $\zeta/s \approx 0.06$ from our present work, $\zeta/s \approx 0.03$ from Duke group, and $\zeta/s \approx  0.02$ from Trajectum.
In general,
our result on $\zeta/s$ is in good agreement with those from holographic model, SMASH transport model and JETSCAPE, and has minor overlap with that from Duke and Trajectum. The discrepancies between Duke, JETSCAPE and Trajectum mainly come from different treatments of observations and parameters~\cite{Nijs:2020ors} although they use the similar multi-stage methods.
The $\zeta/s$ is believed to
have a peak around $T_{\rm pc}$ and go to zero at sufficiently high temperatures~\cite{Karsch:2007jc,Ryu:2015vwa,JETSCAPE:2020mzn}, while the magnitude of the peak value remains unknown and depends on the model~\cite{Shen:2020gef}.

We note that using higher $p_T$ ranges of experimental data has a minor influence on our results.
For example, using $p_T<2.3$ GeV/$c$ for $\phi$ mesons and $p_T<3.6$ GeV/$c$ for $\Omega$ baryons leads to $T=160\pm 8$ MeV, $4\pi\eta/s = 2.14\pm 0.44$ and $\zeta/s = 0.04\pm 0.03$ at $90\%$ C.L.. As expected, the higher $p_T$ ranges have more data points and put stronger constraints on $\eta/s$ and $\zeta/s$.
On the other hand, the higher $p_T$ ranges will cause larger viscous corrections for $f_0$, and the threshold of applicable $p_T$ is not known precisely.
Since using higher $p_T$ ranges is at risk to violate the applicability of the model, we make a conservative choice and use lower ranges in this work. We look forward to using more precise data from future experimental measurements to reduce the uncertainties.

We now discuss the uncertainties in our analysis. In general, we can group them into three categories: (a) Uncertainties from assumptions made in blastwave parameterization, e.g. the simple ansatz for the flow field and the recombinte hypersurface, and the Navier-Stokes approximation for shear stress. (b) Uncertainties from assumptions made in quark recombination formulism, e.g. the instantaneous hadronization process and simple form for wave function squared of hadrons. (c) Uncertainties from the errors in experimental data and the quality of the Gaussian emulator. 
To quantify uncertainty (a), 
one can compare hydro simulations with blastwave, which is what we did in \cite{Yang:2020oig}. We find the uncertainties ($\approx 0.1/4\pi$) from approximations used in blastwave is rather small compared with uncertainties from Bayesian analysis ($\approx 1/4\pi$).
To give a rough estimate for uncertainty (b), 
we vary the wave function width of mesons and baryons by increasing their values by 33\%, i.e., from the default $\sigma_M=0.3$ and $\sigma_B=0.09$ to $\sigma_M=0.4$ and $\sigma_B=0.12$, and we obtain $\eta/s=2.0/4\pi$ and $\zeta/s=0.07$, which correspond to minor changes, i.e., a decrease of $\eta/s$ by 4\% and a increase of $\zeta/s$ by 8\%. We note that for uncertainty (c), one could use more refined recombination formula, such as formula in \cite{Pu:2018eei} to decrease the uncertainty.
Uncertainty (c)
is provided by the MADAI code and is shown in our final results.
In addition, 
there could be uncertainties from the assumption that $\phi$ and $\Omega$ have weak hadronic interactions. In principle, a quantitative estimate of the uncertainties could be obtained by including hadronic transport simulations after hadronization, but this is beyond the scope of the present work and it would be interesting to pursue in future.

Finally, we would like to mention that in the present work, the quark phase-space distribution is parameterized by a viscous blastwave and the $\eta/s$ and $\zeta/s$ of the QGP is only constrained at crossover.
An alternative and perhaps more realistic way is to take the phase-space distribution of quarks from viscous hydrodynamic simulations, which can include parameterized $\eta/s(T)$ and $\zeta/s(T)$ of the QGP.
By doing this, one may extract information on the temperature dependence of the $\eta/s$ and $\zeta/s$ of the QGP from the $\phi$ and $\Omega$ observables.
Such hybrid approach combining hydrodynamic model and quark recombination has been recently proposed in Ref.\ \cite{Zhao:2020wcd},
and in the future,
we may follow the similar approach to investigate the $\eta/s$ and $\zeta/s$ of the QGP from the $\phi$ and $\Omega$ observables.

\section{Conclusions}

Within the quark recombination model with the quark phase-space distribution parameterized in a viscous blastwave, we have demonstrated that the $\phi$ and $\Omega$ produced in relativistic heavy-ion collisions can be used to constrain
the specific shear viscosity $\eta/s$ and the specific bulk viscosity $\zeta/s$ of the QGP at hadronization.
By preforming Bayesian analyses on the measured transverse-momentum spectra and elliptic flows of $\phi$ and $\Omega$ in Au+Au collisions at $\sqrt{s_{\rm NN}} = 200$~GeV and Pb+Pb collisions at $\sqrt{s_{\rm NN}} = 2.76$~TeV, we obtain $\eta/s=(2.08^{+1.10}_{-1.09})/4\pi$ and $\zeta/s= 0.06^{+0.04}_{-0.04}$ at $90\%$ C.L. for the baryon-free QGP at crossover temperature $T_{\rm pc}\approx 160$~MeV.
Our work suggests that high quality data of $\phi$ and $\Omega$
in heavy-ion collisions at various energies may provide important information on
the temperature and baryon density dependence of QGP's $\eta/s$ and $\zeta/s$.

\section*{acknowledgments}

The authors would like to thank Rainer J. Fries, Guang-You Qin and Yifeng Sun for useful discussions. This work was supported by the National Natural
Science Foundation of China under Grant Nos. 12205182, 12235010 and 11625521, and the National SKA Program of China No. 2020SKA0120300.

\bibliography{ref} 

\end{document}